\documentclass[aps,prl,
  twocolumn,
]{revtex4-2}
\usepackage{amsmath,amssymb,amsfonts,amsthm}
\usepackage{bm,mathrsfs,mathtools}
\usepackage{dsfont}
\usepackage{textcomp}
\usepackage{upgreek,textgreek}
\usepackage{physics-patch}
\usepackage{hyperref}
\usepackage{orcidlink}
\usepackage{graphicx}
\usepackage{float}
\usepackage{siunitx}
\usepackage{booktabs}
\usepackage{color}
\usepackage[dvipsnames]{xcolor}
\usepackage{soul}
\usepackage{chngcntr} 

\begin{document}
\raggedbottom
\title{Ionization-Induced Electrostatic Hose Instability in Electron-Beam-Sustained Plasmas}

\author{Jia-Hong Chen}
\affiliation{Sino-French Institute of Nuclear Engineering and Technology, Sun Yat-Sen University, Zhuhai 519082, China}

\author{Yi Yu}%
\affiliation{Sino-French Institute of Nuclear Engineering and Technology, Sun Yat-Sen University, Zhuhai 519082, China}

\author{Jian Chen\,\orcidlink{0000-0001-9807-489X}}%
\email[Contact author:~]{chenjian5@mail.sysu.edu.cn}
\affiliation{Sino-French Institute of Nuclear Engineering and Technology, Sun Yat-Sen University, Zhuhai 519082, China}

\author{Zhi-Bin Wang\,\orcidlink{0000-0002-6812-7855}}%
\email[Contact author.~]{wangzhb8@sysu.edu.cn}
\affiliation{Sino-French Institute of Nuclear Engineering and Technology, Sun Yat-Sen University, Zhuhai 519082, China}

\begin{abstract}
We report the discovery of a previously unrecognized electrostatic hose instability in electron-beam-sustained plasmas, driven by the coupling between the electron beam centroid and the plasma generated via the beam-impact ionization. Unlike the conventional hose instability of relativistic beams propagating in underdense plasmas, this instability requires only ionization-capable electron beams readily produced by common emission processes and sheath acceleration, indicating broad relevance across various discharges. A linear theory is developed to predict the hosing frequency and growth rate, and particle-in-cell/Monte Carlo simulations confirm both the onset of instability and the theoretical predictions.
\end{abstract}
\maketitle

\emph{Introduction.---}
Electron beam–plasma interactions occur in various plasma systems, including low temperature plasma devices~\cite{zhao2019observation,fu2020high,oks1999development,burdovitsin2022plasma,li2013using,levko2016influence,rauf2023particle,rauf2017three,cao2023characterization,chen2025particle}, plasma wakefield acceleration setups~\cite{chen1985acceleration,litos2014high,rosenzweig2005effects,huang2007hosing,mehrling2017mitigation}, and space plasmas~\cite{reid2014review,cairns1999strong,akbari2021langmuir,mcfadden1986high}. These interactions can excite a variety of plasma instabilities~\cite{kaganovich2016band,sakawa1992nonlinear,shustin2021beam} and give rise to phenomena such as nonlinear wave excitation~\cite{drummond1970nonlinear,sun2022physical}, plasma heating~\cite{thode1973two,greenspan1980plasma,kumar2015electron}, turbulence~\cite{sun2022electron,kontar2002nonlinear}, and self-organization~\cite{chen2025three,tyushev2025mode}, thereby strongly influencing both the beam dynamics and the plasma properties~\cite{meger2001beam,fernsler1998production,petrova2024one}. 

Electron hose instability is a well-known transverse beam-plasma instability characterized by the growth of transverse oscillations of the beam centroid. As an intense electron beam propagates through a plasma, it expels plasma electrons and leaves behind an ion channel. A small transverse displacement of the beam centroid then generates an asymmetric wakefield that further deflects trailing beam slices, leading to the characteristic hosing motion of the beam~\cite{davidson2004collective,davidson1980coupled,whittum1991electron,lampe1993electron}. 

Previous studies have primarily focused on intense electron beams propagating through pre-existing underdense plasmas~\cite{lampe1993electron,whittum1991electron,uhm2003effects,uhm2005theory}, motivated largely by applications in plasma-based accelerators. However in many low temperature plasma discharges, the plasma is sustained by the electrons generated through different emission processes. Examples include microdischarges driven by field emission electrons~\cite{li2013using,rumbach2014experimental}, hollow cathode discharges sustained by thermionic emission electrons~\cite{taunay2022physics,goebel2021plasma}, and multipactor-induced discharges sustained by secondary-emission electrons~\cite{wen2022higher,jin2024oblique}. These emitted electrons are accelerated through the sheath, forming an electron beam. Unlike the intense beams expelling plasma electrons in conventional hose instability, these beams generate plasma through ionization. As a result, beam-impact ionization plays an important role in determining the plasma properties~\cite{rauf2023particle,rauf2017three,cao2023characterization,chen2025particle}. Though the influence of ionization on hose instability has previously been considered in the context of plasma wakefield accelerators, where tunnel ionization is induced by the extremely strong electric fields associated with the blowout regime~\cite{deng2006hose}, such mechanisms are not applicable to low temperature plasmas. As a result, whether hose instability can develop in electron-beam-sustained gas discharges and how it may affect plasma properties remain open questions. 

In this Letter, we identify an distinct electrostatic hose instability driven by beam-impact ionization in electron-beam-sustained discharges. Similar to the conventional electron hose instability, the newly identified instability also manifests itself as a growing transverse oscillation of the beam centroid. However, unlike conventional hose instability requiring ultra-intense beams capable of expelling most plasma electrons and forming an ion-focusing channel, the instability reported here naturally arises due to the coupling of the beam and the plasma it generates. It can be triggered by ionization-capable electron beams readily produced through common emission processes and sheath acceleration, making it broadly relevant to low-temperature plasma discharges. We develop a linear theory to predict the hosing frequency and instability growth rate, and validate the theoretical predictions using first-principles particle-in-cell/Monte Carlo collision (PIC/MCC) simulations. We also demonstrate that the instability can lead to beam breakup and generate intense oscillatory particle and energy fluxes to the sidewalls, potentially degrading the low temperature plasma device performance in practical applications. 

\begin{figure}[htbp]
\centering
\includegraphics[width=0.5\textwidth,height=\textheight,keepaspectratio]{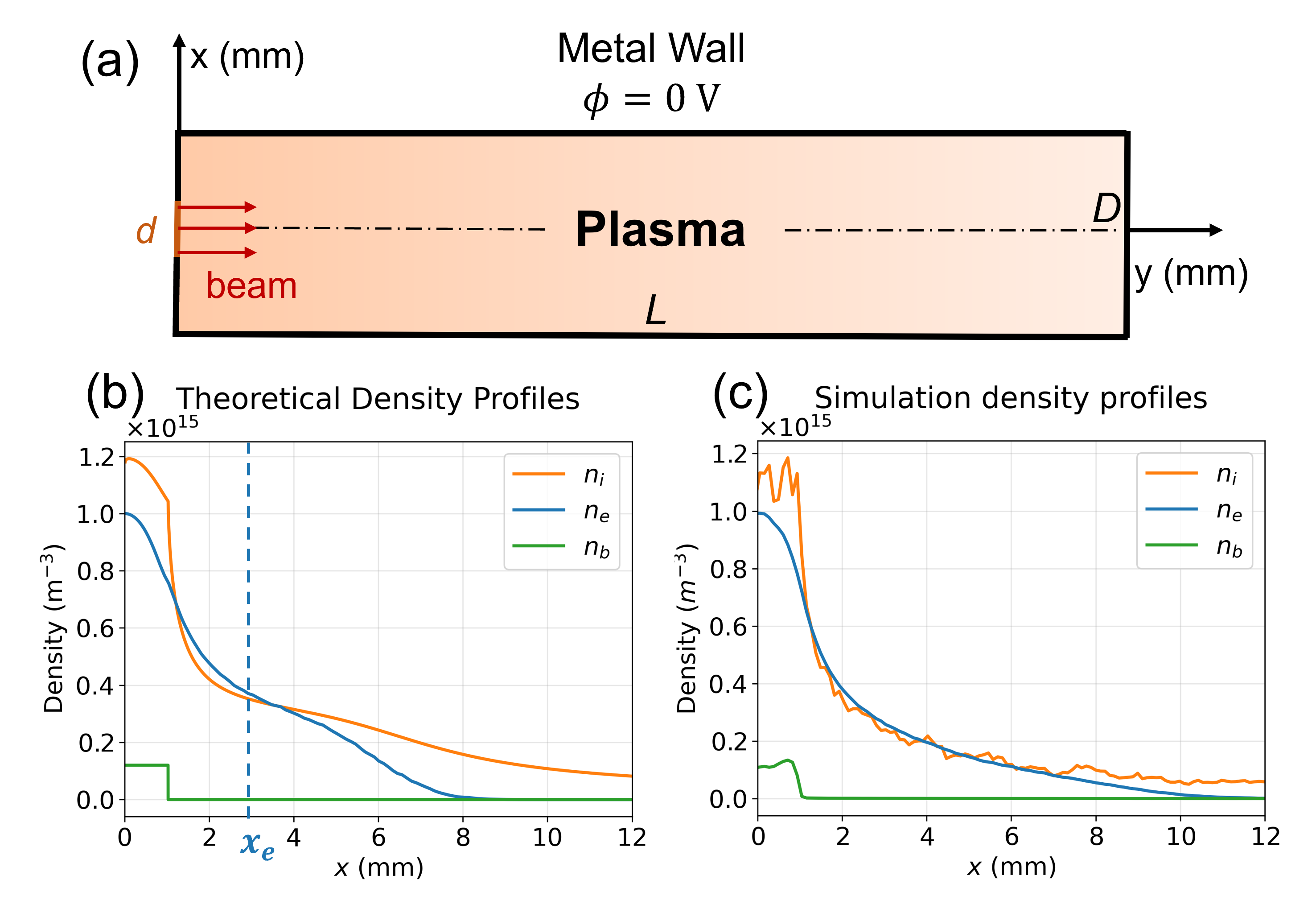}
\caption{\label{fig1}(a) Schematic of the simulation for the electron-beam-sustained discharge. (b) Density profiles obtained by solving Eq.~(\ref{eq:1}) using the parameters listed in TABLE I. (c) Density profiles from PIC simulation along $y=60$ mm at $P$ = 2 mTorr. The orange, blue, and green curves denote the ion, plasma electron, and beam electron densities, respectively. The blue dashed line in (b) marks the edge of the high plasma electron density region, defined by the $1/\rm e$ of its central value.}
\end{figure}

\emph{Transverse profiles of electron-beam-sustained plasmas.---}
Before deriving the hose instability, we need to first calculate the transverse profiles of electron-beam-sustained plasmas. We consider a nonrelativistic electron beam propagating through a background neutral gas without magnetic confinement [Fig.~\ref{fig1}(a)]. As the electron beam ionizes the gas, a plasma is generated. In such an unmagnetized plasma, ions are radially accelerated by the transverse electric field, while the lighter electrons follow a Boltzmann distribution~\cite{dunn1964static,self1963exact,halsted1966electrostatic}. As a result, the transverse profiles satisfy
\begin{equation}
\begin{split}
\frac{d^2 \phi}{dx^2}(x)
&=
-4\pi e n_{e,c}
\left\{
\int_0^x
\frac{\nu_in_b(\xi) \, d\xi}
{n_{e,c}\left(2e/m_i\right)^{1/2}\left[\phi(\xi)-\phi(x)\right]^{1/2}}
\right. \\
&\qquad\left.
-
\exp\!\left(\frac{e\phi(x)}{kT_e}\right)
-
\frac{n_b(x)}{n_{e,c}}
\right\},
\end{split}
\label{eq:1}
\end{equation}
where $\phi$ is the transverse electrostatic potential, $x$ is the transverse coordinate measured from the beam axis, $n_{e,c}$ is the plasma electron density in the center of the beam, $n_b$ is the beam electron density, $T_e$ is the plasma electron temperature, $k$ is the Boltzmann constant and $m_i$ is the ion mass. 
Here, $\nu_i = v_b n_{\mathrm{gas}} \sigma_{\mathrm{ion}}$ is the ionization frequency, with $v_b$ the beam velocity, $n_{\mathrm{gas}}$ the neutral-gas density, and $\sigma_{\mathrm{ion}}$ the ionization cross section at the beam energy. 

Given values of $n_{e,c}$, $n_b$ and $T_e$, Eq.~(\ref{eq:1}) can be solved numerically to obtain the corresponding density and potential profiles. In principle, the evolution of electron beam can also be determined by coupling Eq.~(\ref{eq:1}) to the longitudinal beam dynamics and such a treatment can predict electrostatic beam pinching observed in experiments~\cite{halsted1966electrostatic}. Since this effect is not essential to the onset of the hose instability, it is not presented here.

We performed first-principles particle-in-cell/Monte Carlo simulations to validate Eq.~(\ref{eq:1}). As illustrated in Fig.~\ref{fig1}(a), the simulations employ a two-dimensional Cartesian geometry with dimensions $D = 26~\mathrm{mm}$ and $L = 420~\mathrm{mm}$. A monoenergetic electron beam is injected along the $y$ direction through a $d = 2~\mathrm{mm}$ wide aperture located at the center of the $x$ boundary, with a beam energy of $400~\mathrm{eV}$ and a beam current density of $250~\mathrm{A/m^2}$. These parameters are chosen to match the experimental conditions in Ref.~\cite{halsted1966electrostatic}. More details of the simulations are provided in Appendix~A of Supplemental Material(SM)~\cite{SM}. 

Using the parameters listed in Table I, Eq.~(\ref{eq:1}) predicts the transverse potential and density profiles as shown in Fig.~\ref{fig1}(b). Both profiles exhibit a pronounced hump structure due to the strong ionization within the beam channel. The simulation results [Fig.~\ref{fig1}(c)] confirm the formation of this structure and show agreement with the theoretical predictions. Some discrepancies appear in the peripheral region ($x \gtrsim 8$ mm), likely due to wall effects. As discussed in Appendix~B of SM~\cite{SM}, these discrepancies have a negligible influence on the onset of hose instability.

\begin{table}[htbp]
  \centering
  \caption{Physical parameters of the PIC simulation.\label{tab:hose_frequencies}}
  \begin{tabular}{l l l}
    \hline
    Property & Symbol & Value \\
    \hline
    Electron beam energy & $E_b$(eV) & 400 \\
    Gas pressure & $P$ (mTorr) & 2.0 \\
    Ionization frequency & $\nu_i$(s$^{-1}$) &  $1.2\times10^7$ \\
    Central plasma electron density & $n_{e,c}$(m$^{-3}$) &  $1.0\times10^{15}$ \\
    Plasma electron temperature & $T_{e}$(eV) &  From simulation \\
    Beam electron density & $n_{b}$(m$^{-3}$) &  $1.2\times10^{14}$ \\
    Beam radius & $x_{b}$(mm) &  1.0 \\
    \hline
  \end{tabular}
\end{table}

\emph{Ionization-induced hose instability.---}
Having obtained the transverse profiles, we next examine their stability to transverse hose perturbations. We introduce small transverse displacements to the beam electron and plasma electron channel (denoted by $X_b$ and $X_e$, respectively)~\cite{davidson2004collective,davidson1980coupled,whittum1991electron,lampe1993electron}. For the beam electrons, a rigid transverse displacement produces a density perturbation at the beam edge
\begin{equation}
n_{b1}(x) = n_{b0} X_b \delta(x-b). \label{eq:nb1}
\end{equation}

For the plasma electrons, a transverse displacement $X_e$ yields the density perturbation
\begin{equation}
n_{e1}(x) = -\,\frac{e n_e^{(0)}(x)}{kT_e}\,\frac{d\phi_0}{dx}\,X_e .
\label{eq:ne1}
\end{equation}
For generality, we define the edge of the plasma-electron channel as the position $x=x_e$ where the equilibrium electron density decreases to $1/\mathrm{e}$ of its central value (see blue dashed line in Fig.~\ref{fig1}(b)). 

Besides the electron response, the response of ionization-generated ions is also crucial to the development of the new instability, which is the primary difference from the conventional hose type. Since the ions are produced locally through the beam-impact ionization, their density perturbation is assumed to follow the beam displacement $X_b$. As detailed in Appendix~C of SM~\cite{SM}, the ion density perturbation can be written as $n_{i1}(x) = S_i(x,\omega) X_b$, where
\begin{equation}
S_i(x,\omega) =
\frac{\bigl[1-\exp(-2\pi\nu_i/|\omega_r|)\bigr]
\nu_i n_{b0} H(x-b)}
{\sqrt{2e/m_i}\,\bigl[\phi_0(b)-\phi_0(x)\bigr]^{1/2}}.
\label{eq:ni1}
\end{equation}
The factor [$1-\exp(-2\pi\nu_i/|\omega_r|)$] represents the probability that ionization occurs within one hosing period. In the limit of $\nu_i \ll |\omega_r|$, this factor approaches zero, recovering the pure electron hose case.

Inserting the density perturbations (\ref{eq:nb1})--(\ref{eq:ne1}) into the linearized Poisson equation
\begin{equation}
\frac{d^2\phi_1}{dx^2} = 4\pi e\bigl(n_{e1} + n_{b1} - n_{i1}\bigr),
\end{equation}
the potential perturbation can be written as $\phi_1(x,t)=X_b(t)\psi_b(x)+X_e(t)\psi_e(x)$~\cite{whittum1991electron,lampe1993electron},
where the shape functions $\psi_b$ and $\psi_e$ satisfy
\begin{equation}
    \begin{cases}
        \psi''_b(x) = -4\pi e\bigl[S_i(x,\omega)-n_{b0}\delta(x-b)\bigr] \\[2pt]
        \psi''_e(x) = -4\pi e\,\dfrac{e n_e^{(0)}(x)}{kT_e}\,\phi'_0(x)
    \end{cases}.
\end{equation}
These equations can be solved numerically with the Dirichlet conditions.

For plasma electrons, the transverse displacement follows
\begin{equation}
m_e
\frac{d^2X_e}{dt^2}
=
e
\frac{\partial \phi_1}{\partial x}
|_{x_e}.
\end{equation}
While the displacement for electron beam channel follows
\begin{equation}
m_e
\left(
\frac{d^2X_b}{dt^2}
+
\nu_b\frac{dX_b}{dt}
\right)
=
e
\left.
\frac{\partial \phi_1}{\partial x}
\right|_{b}.
\end{equation}

Therefore, the transverse dynamics can be reduced to two coupled oscillator equations
\begin{equation}
\begin{cases}
\partial_t^2{X}_b + \nu_b\partial_t{X}_b + \Omega_{bb}^2 X_b + \Omega_{be}^2 X_e &= 0 \\
\partial_t^2{X}_e + \Omega_{ee}^2 X_e + \Omega_{eb}^2 X_b &= 0
\end{cases},
\end{equation}
where the coupling constants $\Omega_{bb}, \Omega_{be}, \Omega_{eb}, \Omega_{ee}$ are defined as
\begin{equation}
\begin{cases}
\Omega_{bb}^2=-\frac{e}{m_e}\psi_b'(b), \quad &\Omega_{be}^2=-\frac{e}{m_e}\psi_e'(b)  \\[2pt]
\Omega_{eb}^2=-\frac{e}{m_e}\psi_b'(x_e), \quad &\Omega_{ee}^2=-\frac{e}{m_e}\psi_e'(x_e)
\end{cases}.
\end{equation}

Assuming normal‑mode solutions $X_b, X_e \propto e^{-i\omega t}$, we obtain the dispersion relation
\begin{equation}
\begin{split}
\omega^4
&+
i\nu_b\omega^3
-
\left(
\Omega_{bb}^2+\Omega_{ee}^2
\right)\omega^2\\
&-
i\nu_b\Omega_{ee}^2\omega
+
\Omega_{bb}^2\Omega_{ee}^2
-
\Omega_{be}^2\Omega_{eb}^2
=
0
\end{split},
\label{eq:dispersion}
\end{equation}
with $\omega = \omega_r + i\gamma$. The hose instability corresponds to a positive growth rate, $\gamma > 0$. Eq.~(\ref{eq:dispersion}) can be solved using the transverse profiles calculated from Eq.~(\ref{eq:1}).

\emph{Simulation results and comparison with theory.---}
We performed two‑dimensional PIC/MCC simulations to verify the existence of the hose instability and to benchmark the linear theory. 

\begin{figure*}[htbp]
\centering
\includegraphics[width=\textwidth,height=\textheight,keepaspectratio]{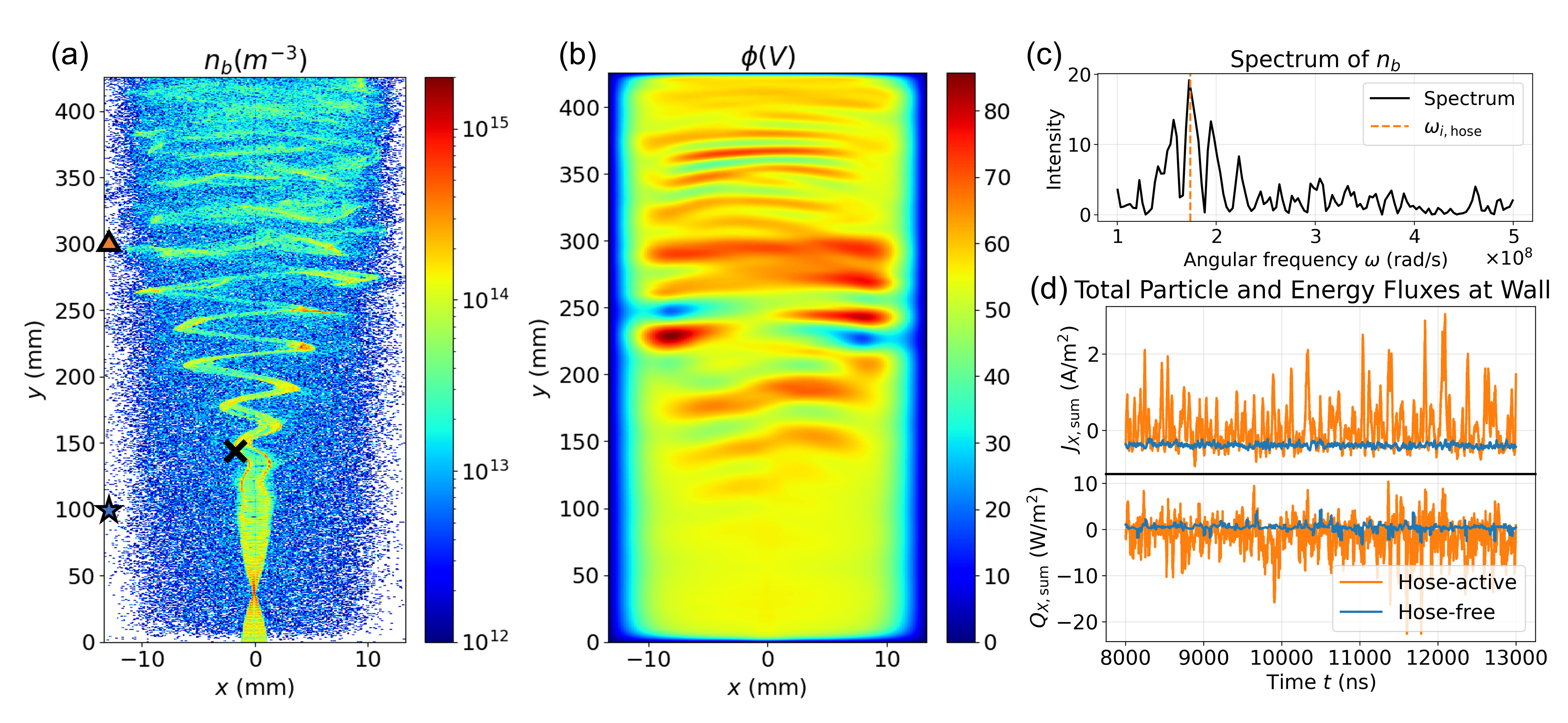}
\caption{\label{fig:simulation} 
(a) and (b) are 2D profiles of beam electron density and electrostatic potential, respectively. The symbols cross, star and triangle mark the probe positions: the cross indicates the probe used for the spectrum in (c), while star and triangle indicate the probes located in the hose‑free and hose‑active regions, respectively, from which the total particle flux $J_{\rm x,sum}$ and total energy flux $Q_{\rm x,sum}$ are recorded.
(c) Fourier spectrum of the beam electron density at cross probe. The orange dashed and blue dashed lines are theoretical ionization‑induced hosing frequency.
(d) Total particle and energy fluxes collected at the wall for the two positions (the star at $y=100$ mm and the triangle at $y=300$ mm), illustrating the contrast between the hose‑active and hose‑free regions. The operating conditions are $P=2$ mTorr and $E_b=400$ eV.}
\end{figure*}

Figs.~\ref{fig:simulation}(a) and (b) show the beam electron density and electrostatic potential distributions at $t$=11 $\mu\mathrm{s}$ for the case with parameters listed in Table ~I. An electrostatic pinch is observed near $y = 40$ mm where the beam radius visibly contracts. This pinch has been reported in experiments~\cite{halsted1966electrostatic} and is a natural consequence of the excess positive charge within the beam channel, as shown in Fig.~\ref{fig1}(c). A growing transverse oscillation of the beam centroid becomes apparent beyond $y = 150$ mm in the beam density profile [Fig.~\ref{fig:simulation}(a)], signaling the onset of a hose instability. The electrostatic potential [Fig.~\ref{fig:simulation}(b)] exhibits a corresponding transverse wobble that oscillates in phase with the beam displacement. As the instability develops, its amplitude increases with propagation distance. In the far downstream region ($y \gtrsim 300$ mm), the beam loses its confined structure and breaks up into a sequence of wavefront-like patterns, indicating the strong impact of the hose instability on the beam dynamics. 

To determine the hosing frequency, we placed a probe near the beam edge (marked by $\times$ in Fig.~\ref{fig:simulation}(a), located at $x=-1$ mm and $y=140$ mm) and recorded the temporal evolution of the beam density. The resulting Fourier spectrum is shown in Fig.~\ref{fig:simulation}(c). Several pronounced peaks appear in the range $1.3\times10^8$–$2.3\times10^8$ s$^{-1}$, with the strongest peak located at $\omega_{i,\rm hose} \approx 1.73 \times 10^8$ s$^{-1}$. The existence of multiple peaks is likely associated with longitudinal electrostatic pinching, which modulates the beam density profile and consequently the local ionization rate. 

To compare with simulation, we solve the dispersion relation Eq.~(\ref{eq:dispersion}) using the transverse profiles in Fig.~\ref{fig1}(b) together with the parameters in Table~I. Two unstable roots are obtained. The first root has an angular frequency $\omega_{i,\rm hose} \approx 1.74 \times 10^8$ s$^{-1}$ and a growth rate of $8.38 \times 10^8$ s$^{-1}$, in good agreement with the dominant peak observed in the simulation [see Fig.~\ref{fig:simulation}(c)]. Since this frequency is comparable to the ionization frequency, we identify the mode as the ionization-induced hose instability. The second unstable root has a higher frequency $\omega = 5.29 \times 10^8$ s$^{-1}$ and a smaller growth rate $5.41 \times 10^8$ s$^{-1}$, corresponding to the the pure electron-hose mode when ionization effects become negligible. This mode is somewhat not seen in the simulated spectrum, likely due to the electron damping. We will leave the nonlinear saturation of these unstable modes for future work.

To further validate the theory, we performed additional simulations with different ionization frequencies. The theoretical predictions for the fastest-growing mode and the simulation results are summarized in Table~II. Good agreement is obtained over the parameter range considered. Furthermore, the hosing frequency scales approximately linearly with the ionization frequency $\nu_i$, providing strong evidence that ionization is the underlying drive of the instability.

Beyond causing beam breakup, the ionization-induced hose instability also strongly modulates the particle and energy fluxes to the sidewalls. As shown in Fig.~\ref{fig:simulation}(d), once the instability develops, the wall fluxes exhibit intense high-frequency bursts. Such fluctuations may degrade the efficiency of plasma processing and material treatment based on electron-beam-sustained discharges, highlighting the need for instability mitigation.

\begin{table}[htbp]
  \centering
  \caption{Comparison of theoretical and simulated hosing frequencies
           for different pressures at $E_b=400$ eV.
           $\omega_{i,\rm hose}^{\rm theo}$ and $\omega_{i,\rm hose}^{\rm sim}$ denote the theoretical and the simulation ionization‑induced hosing frequency, respectively.\label{tab:hose_frequencies}}
  \begin{tabular}{c c c c}
    \hline
    $P$ (mTorr) &  $\nu_i$ (Hz) & 
    $\omega_{i,\rm hose}^{\rm theo}$ (s$^{-1}$) &
    $\omega_{i,\rm hose}^{\rm sim}$ (s$^{-1}$)\\
    \hline
    5 & $3.0\times10^{7}$ &  $3.59 \times 10^8$ & $4.19 \times 10^8$ \\
    2 & $1.2\times10^{7}$ &  $1.74 \times 10^8$ & $1.73 \times 10^8$ \\
    1 & $6\times10^{6}$ &  $7.62 \times 10^7$ & $8.48 \times 10^7$ \\
    0.5 & $3\times10^{6}$ &  $3.66 \times 10^7$ & $4.11 \times 10^7$ \\
    \hline
  \end{tabular}
\end{table}

\emph{Summary.---}
We report a novel ionization-induced electrostatic hose instability in electron-beam-sustained discharges. Unlike the conventional electron hose instability that relies on an ultra-intense beam capable of expelling most plasma electrons, the instability identified here arises naturally from the coupling between the beam and the plasma generated through beam-impact ionization. Therefore, it can occur in many low temperature plasma discharges sustained by ionization-capable electron beams. 

We develop a linear theory to predict the hosing frequency and growth rate of this instability and validate the theory using two-dimensional PIC/MCC simulations. Both the linear theory and simulations show that the hosing frequency is closely linked to the ionization frequency, confirming its ionization-driven nature. The simulations also demonstrate that the instability leads to beam breakup and induces intense, high-frequency particle and energy fluxes to the wall, highlighting its potential impact on discharge performance and the need for effective instability mitigation strategies. 

\emph{Acknowledgments.---}
We are grateful to Dr. Igor D. Kaganovich and Dr. Alexander V. Khrabrov for fruitful discussions. This work was supported by the National Natural Science Foundation of China (Grant No. 12305223) . 

\bibliography{apssamp.bib}

\clearpage

\appendix

\section*{Appendix A: Simulation model}
\setcounter{figure}{0}
\renewcommand{\thefigure}{A\arabic{figure}}

A benchmarked 2D3V electrostatic PIC/MCC software EDIPIC-2D~\cite{villafana20212d,charoy20192d} is employed to model the propagation of nonrelativistic electron beams in a background gas. As shown in Fig.~\ref{fig1}(a), the simulation domain is a two-dimensional Cartesian geometry with dimensions $D = 26~\mathrm{mm}$ and $L = 420~\mathrm{mm}$. A monoenergetic electron beam is injected along the $y$ direction through a $d = 2~\mathrm{mm}$ wide aperture located at the center of the $x$ boundary, with a beam energy of $400~\mathrm{eV}$ and thus a current density of $250~\mathrm{A/m^2}$. All other boundaries are grounded metal walls. Argon is used as the background gas, and the effect of ionization on beam propagation and plasma response is examined under different gas pressures from 0.5 to 5 mTorr at $T=300$ K. The setup of these parameters follows that of the experiment in Ref.~\cite{halsted1966electrostatic}.

Electron--neutral collisions, including elastic scattering, inelastic excitation, and ionization, as well as ion--neutral resonant charge-exchange collisions, are included through the null-collision MCC algorithm~\cite{vahedi1995monte}. The numerical resolution is set to $\Delta x = 111~\mu\mathrm{m}$ and $\Delta t = 15.6~\mathrm{ps}$, sufficient to resolve the Debye length $\lambda_D \approx 235~\mu\mathrm{m}$ and the electron plasma frequency $\omega_{pe} \approx 1.6 \times 10^9~\mathrm{Hz}$. The simulations are initialized without plasma. The particle weight is $1.23\times10^5$. Each case is advanced to a quasi-steady state for more than $12~\mu\mathrm{s}$. The simulations are performed on the Sugon supercomputer using 96 CPU cores, with a typical runtime of about two days for each case.

\begin{figure*}
\centering
\includegraphics[width=\textwidth,height=0.8\textheight,keepaspectratio]{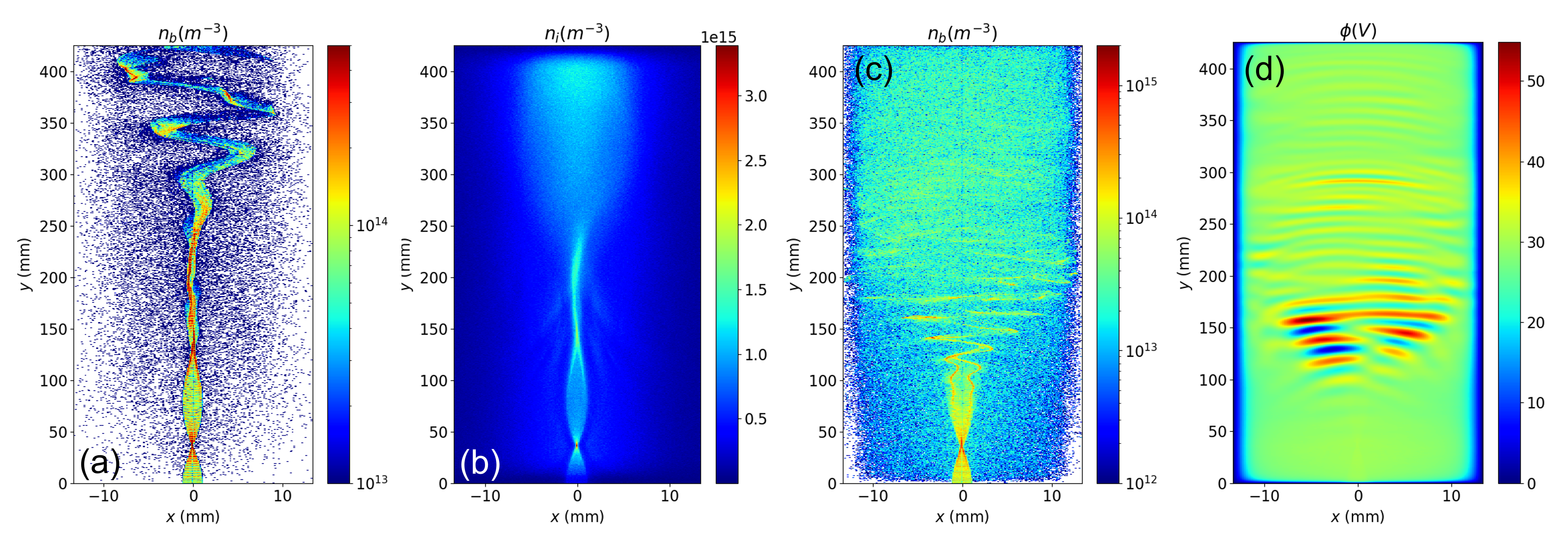}
\caption{\label{fig4}(a) and (b) are 2D spatial distributions of beam electron and ion at $P=$ 1 mTorr. (c) and (d) are 2D spatial distributions of beam electron and potential at $P=$ 5 mTorr.}
\end{figure*}

Figs. ~\ref{fig4}(a)--~\ref{fig4}(b) show the two-dimensional distributions of beam electron or ion at different pressure. In both cases, the electrostatic pinch and the hose instability are clearly observed.
At $P = 1$ mTorr, the ion density profile [Fig.~\ref{fig4}(c)] displays a clear hose‑like transverse oscillation in the region $150 < y < 250$ mm, synchronized with the beam centroid motion. This directly confirms that the ionization‑driven hose instability is imprinted on the ion distribution.

At the higher pressure $P = 5$ mTorr [Fig.~\ref{fig4}(c)], the ionization frequency is larger, leading to a higher hosing growth rate. As a result, the beam electrons are disrupted more rapidly. In the corresponding potential distribution [Fig.~\ref{fig4}(d)], a periodic side‑to‑side oscillation of the potential is clearly observed during the growth phase.

\section*{Appendix B: Nonuniform distribution of Transverse Plasma Electron Temperature}\label{Appendix_1}

\setcounter{figure}{0}
\renewcommand{\thefigure}{B\arabic{figure}}

\begin{figure}[htbp]
\centering
\includegraphics[width=1\columnwidth,height=1\textheight,keepaspectratio]{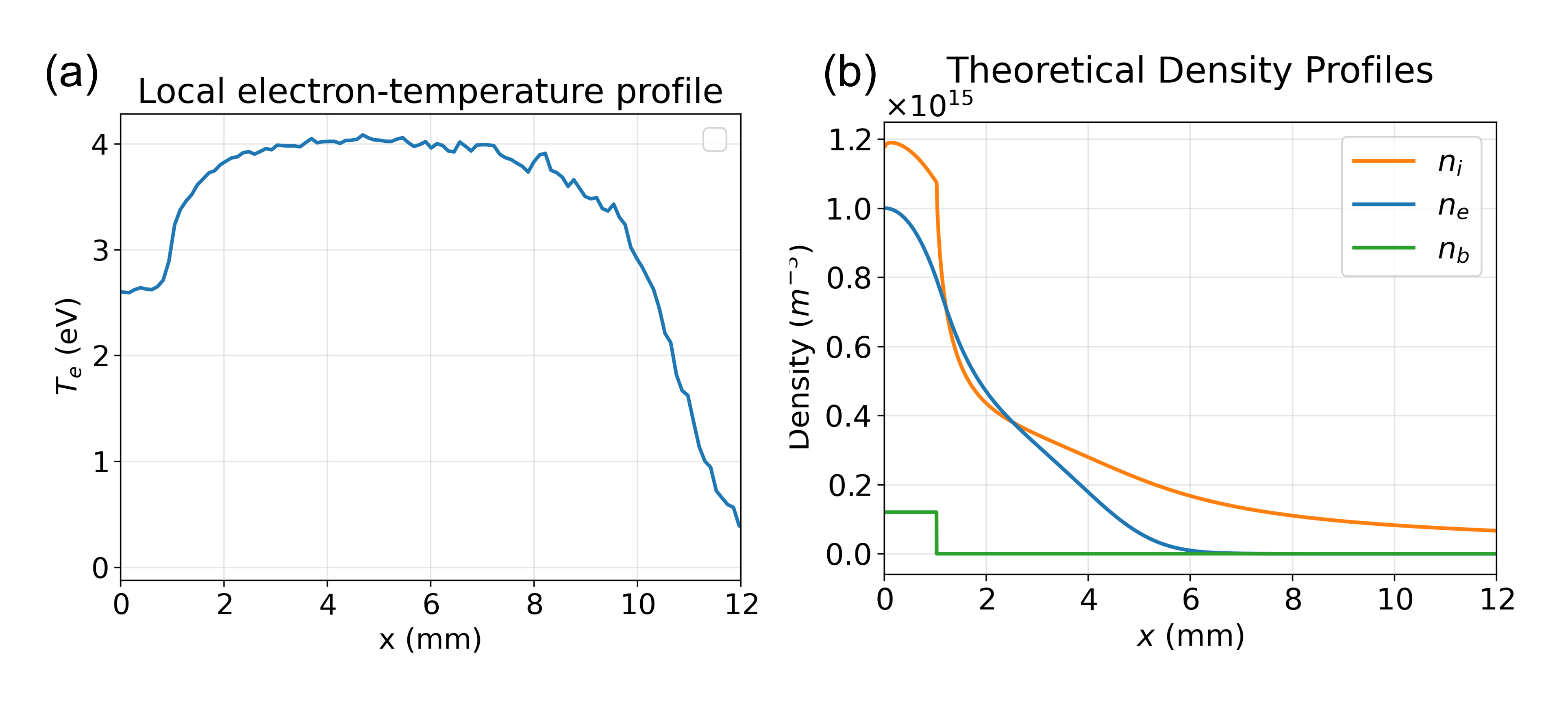}
\caption{\label{fig:appendix_temp}
(a) Electron temperature profile $T_e(x)$ extracted from the PIC simulation.
(b) Density profiles obtained with a fixed plasma electron temperature $T_e = 3.5$ eV. The operating conditions are $P=2$ mTorr and $E_b=400$ eV.}
\end{figure}

Fig.~\ref{fig:appendix_temp}(a) shows the electron temperature profile $T_e(x)$ extracted from the PIC simulation. 
Three distinct regions can be identified: inside the beam channel the temperature is approximately $2.7$ eV; outside the beam ($x \gtrsim 3$ mm) it rises to about $4.1$ eV; and near the wall ($x \gtrsim 10$ mm) it drops sharply to nearly $0$ eV. This non‑uniform temperature motivates the use of a local temperature profile in the Boltzmann relation when computing the transverse profiles.

Fig.~\ref{fig:appendix_temp}(b) compares the plasma electron density profile obtained with a fixed temperature $T_e = 3.5$ eV to that obtained with the local temperature profile from (a). 
The fixed‑temperature calculation still captures the hump structure reasonably well, but the density falls off too steeply in the outer region. This is expected: the actual electron temperature is higher ($\sim 3.5$ eV), and a lower assumed temperature leads to a faster decay of the Boltzmann factor $\exp(e\phi/kT_e)$.

The result using the electron temperature extracted from simulation (used in Fig.~\ref{fig1}(b) of the main text) agrees well with the simulation in the hump region, but also decays faster than the simulation at $x \gtrsim 8$ mm. This residual mismatch may be attributable to wall effects. Near the wall, the rapid drop in electron temperature and the presence of the sheath could cause the local electron population to deviate from a pure Boltzmann distribution. 

Importantly, the characteristic bulk‑electron position $x_e$ (the hump boundary, defined by the $1/\rm e$ density fall‑off) lies around $2$–$4$ mm. In this range the density profiles from the simulation, the local‑temperature calculation, and the fixed‑temperature ($T_e = 3.5$ eV) calculation are all very similar. Consequently, the hosing frequencies derived from these profiles are nearly identical. Using the fixed‑temperature profile we obtain an ionization‑induced hosing frequency $\omega_{i,\rm hose} \approx 1.40 \times 10^8$ s$^{-1}$, which is close to the value obtained with the local‑temperature profile. This confirms that the mismatch in the periphery has negligible impact on the hose instability, and the simpler fixed‑temperature approximation may be sufficient for studying the instability.

\section*{Appendix C: Linear Theory of Hose Instability}\label{Appendix_2}

For a small displacement $X$ and a function $f(x)$,
\begin{equation*}
f(x-X)=f(x)-X\frac{df}{dx}(x)+o(X)
\end{equation*}

In the zeroth-order state, the beam-electron density is taken as a step function,
\begin{equation*}
n_b^{(0)}(x)=n_{b0}H(b-x),
\end{equation*}
where \(b\) is the beam radius. A positive beam displacement \(X_b\) is defined such that the beam boundary moves from \(x=b\) to \(x=b+X_b\). Expanding $H(b-(x-X_b))$ gives,
\begin{equation*}
\begin{split}
H(b-x+X_b)&=H(b-x)+X_b\frac{dH(b-x)}{d(b-x)}+o(X_b)\\
&=H(b-x)+X_b\delta(b-x)+o(X_b)
\end{split}.
\end{equation*}
Thus the first-order beam-electron density perturbation is
\begin{equation*}
\boxed{
n_{b1}(x)=n_{b0}X_b\delta(x-b)
}.
\end{equation*}

The zeroth-order plasma-electron density is assumed to satisfy the Boltzmann relation
\begin{equation*}
n_e^{(0)}(x)
=
n_{e,c}
\exp\left(
\frac{e\phi_0(x)}{kT_e}
\right).
\end{equation*}
Expanding $\exp\left(\frac{e\phi_0(x)}{kT_e}\right)$ gives,
\begin{equation*}
\begin{split}
\exp\left(\frac{e\phi_0(x-X_e)}{kT_e}\right)=&\exp\left(\frac{e\phi_0(x)}{kT_e}\right)\\
&-X_e\frac{e\phi_0'(x)}{kT_e}\exp\left(\frac{e\phi_0(x)}{kT_e}\right)+o(X_e)
\end{split}.
\end{equation*}

Thus the plasma-electron density perturbation in the displacement form becomes
\begin{equation*}
\boxed{
n_{e1}(x)
=
-
\frac{e n_e^{(0)}(x)}{kT_e}
\frac{d\phi_0}{dx}
X_e
}.
\end{equation*}

The zeroth-order ion density is written as
\begin{equation*}
n_i^{(0)}(x)
=
\int_0^x
\frac{
\nu_i n_{b0}H(b-\xi)\,d\xi
}{
\left(2e/m_i\right)^{1/2}
\left[
\phi_0(\xi)-\phi_0(x)
\right]^{1/2}
}.
\end{equation*}
Thus the first-order ion density perturbation can be defined as,
\begin{equation*}
\begin{split}
n_{i1}^{(b)}(x)
&=
\int_0^x
\frac{
\nu_i n_{b1}d\xi
}{
\left(2e/m_i\right)^{1/2}
\left[
\phi_0(\xi)-\phi_0(x)
\right]^{1/2}
}\\
&=
\int_0^x
\frac{
\nu_i n_{b0}X_b\delta(\xi-b)d\xi
}{
\left(2e/m_i\right)^{1/2}
\left[
\phi_0(\xi)-\phi_0(x)
\right]^{1/2}
}. 
\end{split}
\end{equation*}

Since
\begin{equation*}
\int_0^x f(\xi)\delta(\xi-b)\,d\xi
=
f(b)H(x-b),
\end{equation*}
the ion density perturbation due to the displacement of the ionization source is
\begin{equation*}
n_{i1}^{(b)}(x)
=
\frac{
\nu_i n_{b0}X_b H(x-b)
}{
\left(2e/m_i\right)^{1/2}
\left[
\phi_0(b)-\phi_0(x)
\right]^{1/2}
}.
\end{equation*}

However, when the ionization frequency $\nu_i$ is not sufficiently larger than the hosing frequency, the ionization cannot follow the beam displacement instantaneously. Over one oscillation period $T = 2\pi/|\omega_r|$ (with $\omega_r = \operatorname{Re}\omega$), the efficiency with which ionization builds up the density perturbation scales as $1 - e^{-\nu_i T}$. Thus, in the frequency domain the ion density perturbation becomes
\begin{equation*}
{n}_{i1}(x,\omega)
=
\frac{[1 - \exp\left(-2\pi\nu_i/|\omega_r|\right)]
\nu_i n_{b0} H(x-b)
}{
\left(2e/m_i\right)^{1/2}
\left[
\phi_0(b)-\phi_0(x)
\right]^{1/2}
}X_b.
\end{equation*}
In the limit $\nu_i \gg |\omega_r|$ the bracket $[1 - \exp\left(-2\pi\nu_i/|\omega_r|\right)]$ tends to $1$, recovering the instantaneous‑ion result; for $\nu_i \ll |\omega_r|$ the perturbation is strongly suppressed, which accounts for the mismatch between the ionization rate and the hosing frequency.

For later use, we define the ion source response function
\begin{equation*}
S_i(x,\omega)
=
\frac{(1 - \exp\left(-2\pi\nu_i/|\omega_r|\right))
\nu_i n_{b0}H(x-b)
}{
\left(2e/m_i\right)^{1/2}
\left[
\phi_0(b)-\phi_0(x)
\right]^{1/2}
}.
\end{equation*}
Then the ion density perturbation can be written compactly as
\begin{equation*}
\boxed{n_{i1}(x)
=
S_i(x,\omega)X_b}.
\end{equation*}

The first-order Poisson equation is
\begin{equation*}
\frac{d^2\phi_1}{dx^2}
=
4\pi e
\left(
n_{e1}+n_{b1}-n_{i1}
\right).
\end{equation*}

Substituting the displacement forms of \(n_{i1}\), \(n_{e1}\), and \(n_{b1}\), one obtains
\begin{equation*}
\begin{split}
\frac{d^2\phi_1}{dx^2}
=
4\pi e
[&X_b
\left(
n_{b0}\delta(x-b)-S_i(x,\omega)
\right)\\
&-
\frac{e n_e^{(0)}(x)}{kT_e}
\frac{d\phi_0}{dx}
X_e]
\end{split}.
\end{equation*}

Because the equation is linear in $X_b$ and $X_e$, the first-order potential can be decomposed as
\begin{equation*}
\phi_1(x,t)
=
X_b(t)\psi_b(x)
+
X_e(t)\psi_e(x),
\end{equation*}
where $\psi_b$ is the potential response to a unit beam displacement and $\psi_e$ is the potential response to a unit plasma-electron displacement.

The two response functions satisfy
\begin{equation*}
\boxed{
\frac{d^2\psi_b}{dx^2}
=
-4\pi e
\left[
S_i(x,\omega)-n_{b0}\delta(x-b)
\right]
}
\end{equation*}
and
\begin{equation*}
\boxed{
\frac{d^2\psi_e}{dx^2}
=
-4\pi e
\frac{e n_e^{(0)}(x)}{kT_e}
\frac{d\phi_0}{dx}
}.
\end{equation*}

For a half-domain description of a hose-like transverse displacement, the first-order potential is taken to be odd with respect to the beam center. Thus one may impose
\begin{equation*}
\psi_b(0)=0,
\qquad
\psi_e(0)=0.
\end{equation*}

At the outer boundary, a simple grounded condition is used,
\begin{equation*}
\psi_b(L)=0,
\qquad
\psi_e(L)=0.
\end{equation*}

The beam centroid is driven by the first-order electric field at the inner side of the beam boundary. Including a beam-electron collision frequency \(\nu_b\), its equation of motion is written as
\begin{equation*}
m_e
\left(
\frac{d^2X_b}{dt^2}
+
\nu_b\frac{dX_b}{dt}
\right)
=
e
\left.
\frac{\partial \phi_1}{\partial x}
\right|_{b^-}.
\end{equation*}

Substituting
\begin{equation*}
\phi_1(x,t)
=
X_b(t)\psi_b(x)
+
X_e(t)\psi_e(x),
\end{equation*}
gives
\begin{equation*}
m_e
\left(
\frac{d^2X_b}{dt^2}
+
\nu_b\frac{dX_b}{dt}
\right)
=
e
\left[
X_b\psi_b'(b^-)
+
X_e\psi_e'(b^-)
\right].
\end{equation*}

This can be written as
\begin{equation*}
\boxed{
\frac{d^2X_b}{dt^2}
+
\nu_b\frac{dX_b}{dt}
+
\Omega_{bb}^2X_b
+
\Omega_{be}^2X_e
=
0
}
\end{equation*}
with
\begin{equation*}
\boxed{
\Omega_{bb}^2
=
-\frac{e}{m_e}\psi_b'(b^-)
}
\end{equation*}
and
\begin{equation*}
\boxed{
\Omega_{be}^2
=
-\frac{e}{m_e}\psi_e'(b^-)
}.
\end{equation*}

The plasma-electron displacement is assumed to occur around a characteristic position $x_e$, which is chosen from the zeroth-order density profile. The characteristic plasma-electron displacement position is defined by the point where the hump density decays to $1/\mathrm e$ of its central value, i.e.,
\begin{equation*}
\boxed{
n_e^{(0)}(x_e)
=
\frac{1}{\mathrm e}n_e^{(0)}(0)=\frac{n_{e,c}}{\mathrm e}
}.
\end{equation*}

Neglecting collisional damping of the plasma electrons, the plasma-electron displacement satisfies
\begin{equation*}
m_e
\frac{d^2X_e}{dt^2}
=
e
\frac{\partial \phi_1}{\partial x}
|_{x_e}
\end{equation*}

Substituting the response decomposition for \(\phi_1\), one obtains
\begin{equation*}
m_e
\frac{d^2X_e}{dt^2}
=
e
\left[
X_b\psi_b'(x_e)
+
X_e\psi_e'(x_e)
\right].
\end{equation*}

Therefore,
\begin{equation*}
\boxed{
\frac{d^2X_e}{dt^2}
+
\Omega_{ee}^2X_e
+
\Omega_{eb}^2X_b
=
0
}
\end{equation*}
where
\begin{equation*}
\boxed{
\Omega_{eb}^2
=
-\frac{e}{m_e}\psi_b'(x_e)
}
\end{equation*}
and
\begin{equation*}
\boxed{
\Omega_{ee}^2
=
-\frac{e}{m_e}
\psi_e'(x_e)
}.
\end{equation*}

Thus, in the instantaneous-ion-response approximation, the coupled displacement equations are
\begin{equation*}
\boxed{
\begin{split}
\frac{d^2X_b}{dt^2}
+
\nu_b\frac{dX_b}{dt}
+
\Omega_{bb}^2X_b
+
\Omega_{be}^2X_e
&=0,
\\
\frac{d^2X_e}{dt^2}
+
\Omega_{ee}^2X_e
+
\Omega_{eb}^2X_b
&=0.
\end{split}
}
\end{equation*}

Assuming normal-mode solutions of the form
\begin{equation*}
X_b,\;X_e
\propto
\exp(-i\omega t),
\end{equation*}
the coupled equations become
\begin{equation*}
\left(
-\omega^2
-i\nu_b\omega
+
\Omega_{bb}^2
\right)X_b
+
\Omega_{be}^2X_e
=
0,
\end{equation*}
and
\begin{equation*}
\Omega_{eb}^2X_b
+
\left(
-\omega^2
+
\Omega_{ee}^2
\right)X_e
=
0.
\end{equation*}

A nontrivial solution requires the determinant to vanish, giving the dispersion relation
\begin{equation*}
\boxed{
\left(
-\omega^2
-i\nu_b\omega
+
\Omega_{bb}^2
\right)
\left(
-\omega^2
+
\Omega_{ee}^2
\right)
-
\Omega_{be}^2\Omega_{eb}^2
=
0
}
.
\end{equation*}

Equivalently, this can be written as a fourth-order polynomial in \(\omega\),
\begin{equation*}
\boxed{
\begin{split}
\omega^4
&+
i\nu_b\omega^3
-
\left(
\Omega_{bb}^2+\Omega_{ee}^2
\right)\omega^2\\
&-
i\nu_b\Omega_{ee}^2\omega
+
\Omega_{bb}^2\Omega_{ee}^2
-
\Omega_{be}^2\Omega_{eb}^2
=
0
\end{split}
}.
\end{equation*}

With the convention \(\exp(-i\omega t)\), a root
\begin{equation*}
\omega=\omega_r+i\gamma
\end{equation*}
corresponds to an instability when
\begin{equation*}
\gamma>0.
\end{equation*}

\end{document}